\documentclass[12pt]{iopart}

\usepackage{iopams}
\usepackage{graphicx}
\usepackage[numbers,square,sort&compress]{natbib}
\usepackage{color}

\begin{document}

%Title of paper
\title[Long-ranged magnetic proximity effects in noble metal-doped cobalt]{Long-ranged magnetic proximity effects in noble metal-doped cobalt probed with spin-dependent tunnelling}

\author{M~S~Gabureac$^1$\footnote{Present address: ETH Z\"{u}rich, Magnetism and Interface Physics Group, Department of Materials, Schafmattstrasse 30, CH-8093 Z\"{u}rich, Switzerland.}, D~A~MacLaren$^2$, H~Courtois$^3$ and C~H~Marrows$^1$\ead{c.h.marrows@leeds.ac.uk}}

\address{1 School of Physics \& Astronomy, University of Leeds, Leeds LS2 9JT, United Kingdom}

\address{2 SUPA, School of Physics \& Astronomy, University of Glasgow, Glasgow G12 8QQ, United Kingdom}

\address{3 Institut N\'{e}el, CNRS/UJF, 25 Avenue des Martyrs, 38042 Grenoble Cedex 9, France}

\begin{abstract}
  We have inserted non-magnetic layers of Au and Cu into sputtered AlO$_x$-based magnetic tunnel junctions and Meservey-Tedrow junctions in order to study their effect on the tunnelling magnetoresistance (TMR) and spin-polarisation (TSP). When either Au or Cu are inserted into a Co/AlO$_x$ interface, we find the TMR and TSP remain finite and measurable for thicknesses up to several nanometres. High-resolution transmission electron microscopy shows that the Cu and Au interface layers are fully continuous when their thickness exceeds $\sim 3$~nm, implying that spin-polarized carriers penetrate the interface noble metal to distances exceeding this value. A power law model based on exchange scattering is found to fit the data better than a phenomenological exponential decay. The discrepancy between these lengthscales and the much shorter ones reported from X-ray magnetic circular dichroism studies of magnetic proximitisation is ascribed to the fact that our tunnelling transport measurements selectively probe $s$-like electrons close to the Fermi level. When a 0.1~nm thick Cu or Au layer is inserted within the Co, we find that the suppression of TMR and TSP is restored on a lengthscale of $\lesssim 1$~nm, indicating that this is a sufficient quantity of Co to form a fully spin-polarised band structure at the interface with the tunnel barrier.
\end{abstract}


\submitto{\NJP}
% Comment out if separate title page not required
\maketitle

\section{Introduction}

The superconducting proximity effect, where Cooper-paired carriers and hence superconducting order can leak into a normal metal that is in contact with a superconductor, is by now well established \cite{michael_tinkham_introduction_1996}. There is an associated inverse proximity effect, where the same leakage causes a suppression of superconducting order in the superconductor. The analogous magnetic proximity effect, where spin-polarized carriers and hence ferromagnetic order can leak from a ferromagnet into an otherwise non-magnetic material and magnetise it (and any associated inverse effect) has received less attention. The detection of proximity-induced magnetism in a material in contact with a ferromagnet is an experimental challenge using conventional magnetometry, since the weak moment in the proximitised material is swamped by the larger moment of the magnetising material. More specialised methods therefore have to be brought to bear. Transition metal ferromagnets in contact with a noble metal \cite{samant1994,bartolome2008}, a transition metal \cite{huttel2008}, a spin glass \cite{abes2009}, an actinide metal \cite{springell2008}, and a dilute magnetic semiconductor \cite{maccherozzi2008,mark2009,olejnik2010,nie2013} have been studied by X-ray magnetic circular dichroism (XMCD) methods. Those experiments used the fact that XMCD is able to separate the magnetic signal from different chemical elements \cite{stohr1999}.

Another method of discriminating between the two materials is to employ a surface sensitive technique. Whilst XMCD has this property in the total electron yield mode, being sensitive only to the first few nanometres, tunnelling studies offer extreme surface sensitivity, since tunnel currents usually arise from only the last plane of atoms at a surface \cite{meserveytedrow}. The first report of a magnetic tunnel junction (MTJ) was made over 30 years ago by Julliere \cite{julliere}. These structures consist of an ultrathin insulating barrier separating two ferromagnetic electrodes (a so-called FIF configuration). Changing the relative alignment of the magnetic moments in the two layers, for instance by application of a magnetic field, changes the junction resistance: this is the tunnelling magnetoresistance (TMR) effect. Since their discovery there has been a huge upsurge of interest in these structures \cite{baderandparkin} as the improvement of fabrication techniques has led to values of room temperature TMR that are large enough for technological applications \cite{moodera1995,yuasanatmat2004,parkinnatmat2004}. The magnitude of the TMR, defined as the fractional change in junction resistance $\Delta R/R$ on switching the junction from a parallel (P) to an antiparallel (AP) magnetization state, depends on the degree of tunnelling spin polarisation (TSP) of the F electrode materials through the Julliere formula \cite{julliere}, which can be barrier-dependent \cite{slonczewski1989}. Since the TMR depends on the geometrical average of the TSPs of the two electrodes, gaining knowledge of the TSP of the ferromagnet to be studied depends on the fact that the TSP of the other is already known. The TSP can also be measured in a way that is independent of any other material by forming a junction between the ferromagnet in question and a superconductor (a SIF structure) by a means devised by Meservey and Tedrow \cite{tedrowmeservey}, where fully spin-polarised states are generated at the edge of the Bardeen-Cooper-Schrieffer (BCS) gap in the superconductor by Zeeman splitting using the application of a magnetic field. The tunnelling states of the ferromagnet in question can then be projected onto these states.

Here we report tunnelling studies of the influence of the magnetic proximity and inverse proximity effects on both TMR and TSP in the materials systems Co/Cu and Co/Au. Both noble metals have very similar electronic structures to that of the spin-$\uparrow$ majority states in Co, but Au is much heavier than Cu and so has a stronger spin-orbit interaction. We have introduced the Cu and Au layers both at the interface between a Co electrode and an AlO$_x$ barrier to probe the manner in which they are proximitised by Co, and also as a $\delta$-layer, which we can place at a variable distance from that interface, to study the inverse proximitisation of the nearby Co. This second case mirrors a previous study of the giant magnetoresistance (GMR) in Co/Cu/Co spin-valves \cite{marrows2001}. Ferromagnetic interface layers have also been studied in GMR systems \cite{parkin1993,stanley2000}. There are reports of previous experiments on MTJs similar to our interface layer studies. Moodera \textit{et al.} deposited Au layers onto Co electrodes before covering them with plasma-oxidized alumina, and found that the TMR almost vanished for Au layers thicker than about 1 nm, in addition to a weak quantum well oscillation \cite{moodera1999}. Other spacer materials have also been studied, such as Cu \cite{sun1999,leclair2000,yuasa2002,samant2004}, Cr \cite{leclair2001cr}, and Ru \cite{leclair2001,nozaki2004}. A variety of different effects, some pointing to the presence of quantum well states, were found \cite{yuasa2002,leclair2001}. On the other hand, we know of no comparable studies in the tunnelling regime for the $\delta$-layer experiments, beyond our preliminary report on our results with Au \cite{gabureac2008}. We find that TSP can penetrate several nanometres of both Cu and Au, and arises from interactions with the first nanometre or so of the Co. These distances are remarkably long, and certainly long enough that the noble metal layers are able to become continuous well before the TMR and TSP are suppressed, which we demonstrate using transmission electron microscopy. The large-thickness dependence of the TMR and TSP is not easily described using a exponential decay, and so we interpret the data within a modified exchange-scattering model \cite{moodera1989} that predicts a power law behaviour. We attribute the discrepancy with the XMCD-based studies to the fact that the two techniques probe different populations of electrons.

\section{Experimental methods}

\begin{figure}[tbp]
  \begin{indented}\item[]
  \includegraphics[width=10cm]{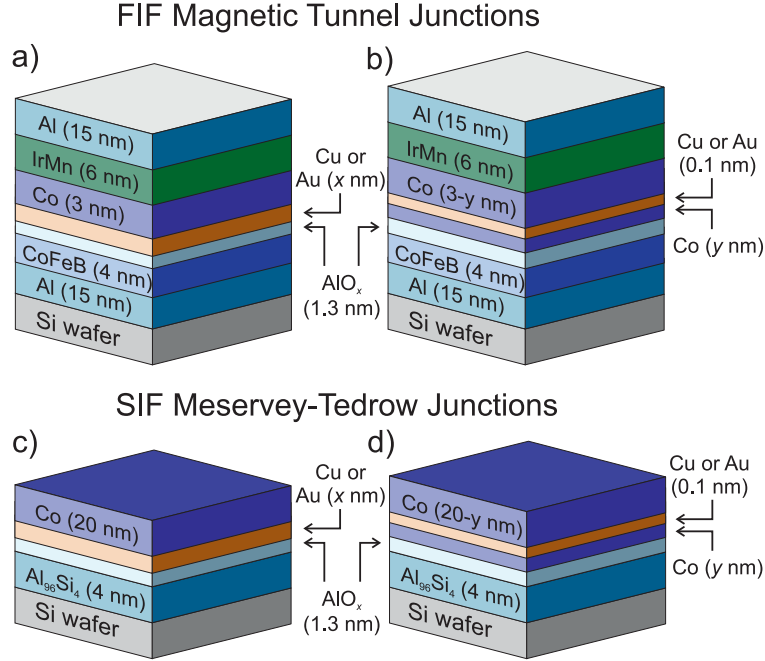}
  \end{indented}
  \caption{Sample structure schematics, showing layer stack sequence deposited on Si wafer substrates. FIF magnetic tunnel junctions with a) interface doping layers and b) $\delta$-doping layers of Cu and Au. SIF Meservey-Tedrow junctions with c) interface doping layers and d) $\delta$-doping layers of Cu and Au. Layer thicknesses are not drawn to scale.} \label{stacks}
\end{figure}

Our junctions were of the cross-strip form, deposited by dc magnetron sputtering through shadow masks, which were changed \textit{in situ} to give an active area of 50 $\mu$m~$\times$~50 $\mu$m. The substrates were pieces of Si wafer with $\sim 100$~nm thermal oxide at the surface. The chamber base pressure was $\sim 2 \times 10^{-8}$~Torr, whilst the working pressure of Ar was 2.5~mTorr. Typical deposition rates, as calibrated on sheet films by X-ray reflectometry, were in the range 0.2--0.3~nm/s for the metal layers, although these are reduced substantially when depositing through the narrow slot of the shadow mask. The expected nominal rate of deposition of the critical Cu and Au layers (see sample description below) is therefore only 0.1 nm/s. The barriers were formed by dc plasma oxidation in 55~mTorr of O$_2$ for 30~s at a power of 100~W. The nominal layer stacks are depicted in figure \ref{stacks}.

Two types of junctions were grown: in each case a Co layer was the object of study. The first were FIF MTJs based on the following layer stack sequence: substrate / Al (15~nm) / Co$_{68}$Fe$_{22}$B$_{10}$ (4~nm) / AlO$_x$ (1.3~nm) / Co (3~nm) / Ir$_{20}$Mn$_{80}$ (6~nm) / Al (15~nm). The IrMn layer pins the Co through the exchange bias effect to control the switching between the P and AP states, which was accomplished by applying small fields to the magnetically soft CoFeB layer.

The second set of junctions were of the SIF type, based on the stacking sequence Al$_{96}$Si$_4$ (4~nm) / AlO$_x$ (1.3~nm) / Co (20~nm). The superconducting Al$_{96}$Si$_4$ layer had a critical temperature, $T_{\rm C}$, of $\sim 3$ K and an in-plane critical field, measured at 1.4~K, exceeding 5~teslas; both quantities were determined from two point resistance measurements of a bottom electrode strip.

The junctions were doped by introducing a thin layer of one of the noble metals Cu or Au. This was done either using a layer of thickness $x$ at the AlO$_x$ barrier/Co interface, to produce \emph{interface-doped} junctions, or at a point within the Co layer a distance $y$ from that interface, to yield \emph{$\delta$-doped} junctions. It is also important to emphasize that the growth protocol used here has been demonstrated to give highly reproducible results: several nominally identical undoped MTJs had TMR ratios and resistance-area products that were the same to within only a few per cent. This means that sample-to-sample variations should not mask any changes due to the introduction of Cu or Au that are above this level. The good tunnelling $I$-$V$ properties of our junctions (shown later), along with X-ray reflectometry of comparable sheet film test samples, confirm that interfacial roughnesses are all well below 1 nm, and our barrier is smooth and pinhole-free.

Cross-sectional samples for electron microscopy were prepared using standard lift-out and low-energy polishing techniques in an FEI Nova Focused Ion Beam (FIB) system. Electron energy loss spectroscopy (EELS) measurements (see below) indicated these cross-section samples to have a typical thickness of 70~nm. Transmission electron microscopy (TEM) was performed on a combination of an FEI Tecnai F20 and a JEOL ARM-cFEG microscope, both operated at 200~kV and equipped with field emission guns and scanning capability. The latter instrument is also equipped with a Gatan Quantum spectrometer with fast shuttering capability that was used for EELS. EELS data sets were acquired in scanning TEM (STEM) mode, using the ‘dual EELS’ technique \cite{scott2008} to collect both the intense zero-loss and weak core-loss spectral features simultaneously. The spectrum imaging technique was used for elemental analysis, whereby an EELS spectrum was acquired at every pixel in a selected region of a STEM image \cite{hunt1991}. Compositions were derived by integrating the EELS spectral intensity of the L$_{2,3}$ excitation edges of Co and Cu and the M$_{4,5}$ edge of Au after removal of a power-law background and assuming standard scattering cross-sections. A number of samples were imaged by TEM and here we will present the analysis of two representative data sets.

The TMR of the FIF MTJs was measured at room temperature by a conventional four-probe dc technique at room temperature: measurements were taken under variable bias, but we only show data here for a rather low 10 mV applied voltage bias in order to measure the TSP close to the Fermi level. All junctions were measured in their as-grown state, as a precaution against any thermally-activated migration of the noble metal layers within the structure during any anneal, for instance along grain boundaries. In no case did the insertion of the Cu or Au layers affect the exchange bias of the Co layer sufficiently that we were unable to determine the full TMR due to the lack of a properly antiparallel state. The differential conductance ($G_{\rm diff} = dI/dV$) as a function of applied bias $V$ of the SIF junctions was measured at 250~mK in a $^3$He refrigerator, under a constant applied field of 2~T supplied by a superconducting magnet. A separate transverse coil was used to null off any small component of field normal to the junction surface due to sample misalignment, which would otherwise give rise to vortices in the superconductor. These transport measurements were performed first, with the TEM specimens subsequently prepared from the active area of the very same samples, for the most direct possible comparison between the two sets of data.

\section{Results \&\ discussion}

\subsection{Interface layer continuity: junction characterisation by TEM}

First we discuss the nature of the inserted Cu and Au layers as determined by TEM, since it is important to answer the question as to whether these thin interface layers of Cu and Au are continuous. If the noble metal layer is not proximitised but is discontinuous, then we would expect that the decrease in TSP reflects the fraction of the interfacial area of the barrier that is not in contact with the Co. In this case the value of $\lambda$ reflects the rate at which pinholes in the noble metal are filled in as the layer thickness $x$ increases. We gave indirect arguments as to why this seems implausible in our previous paper on Au interface layers \cite{gabureac2008}, but expand on these here now that a detailed TEM study has been performed.

On the other hand, if the Au and Cu layers are continuous, then the TMR and Meservey-Tedrow  measurements in the interface-doped samples are sensitive to the degree of spin-polarisation in the carriers at the interface between the noble metal and the AlO$_x$ barrier, this polarisation having arisen through proximity magnetism with the Co. The values of $\lambda$ we obtain would then reflect the lengthscale over which this proximity polarisation decays. X-ray circular dichroism measurements show that there is a detectable moment on the $d$-electrons in Cu adjacent to Co on distances exceeding 1~nm \cite{samant1994,abes2009} in multilayers, and Cu and Au capped Co nanoparticles \cite{bartolome2008}. We should therefore expect $s$-electron polarisation at the Fermi level to at least this distance, with concomitant TMR. Meanwhile, slow muons show long distance polarisation over a few nanometres in Ag in an Fe/Ag/Fe trilayer \cite{luetkens2003}. Furthermore, TMR was observed through 4~nm of Cu by Sun and Freitas \cite{sun1999}, and TSP through several nanometres by Moodera \textit {et al.} \cite{moodera1989}. In the latter case, pinhole contacts between the barrier and ferromagnet were rigourously eliminated by an additional oxidation step.

\begin{figure}[tbp]
  \begin{indented}\item[]
  \includegraphics[width=13cm]{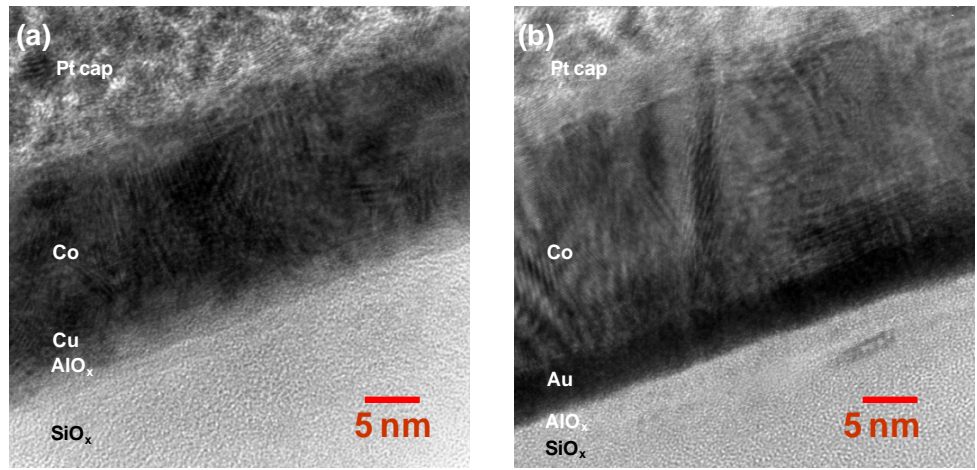}
  \end{indented}
  \caption{Bright field TEM images of typical SIF devices incorporating (a) a nominally 1.6~nm thick Cu and (b) a nominally 3.2~nm thick Au insertion layer. Contrast is dominated by variations in atomic number, so that the Au layer is clearly distinguished as a dark band whilst the Cu layer is harder to discern against the upper Co electrode.}
  \label{tem_sections}
\end{figure}

Figure \ref{tem_sections} shows bright field TEM images of two SIF samples incorporating (a) a Cu and (b) a Au interface layer that are representative of all of the samples studied. (In the FIF junctions, the relevant noble metal and Co layers are grown on top of an identically prepared alumina barrier.) The TEM specimens were prepared from the junctions after the magnetotransport data were acquired. The images have been selected to illustrate a discontinuous and a continuous insertion layer, respectively. The samples have insertion layers of a nominal thickness of 1.6~nm of Cu and 3.2~nm of Au. The images show cross-sections of the SIF stack after capping with a protective Pt layer that was deposited by ion-assisted deposition in the FIB system prior to sample lift-out. Contrast is dominated by variations in atomic number and diffraction, the former causing the thick Co layers to appear dark whilst the latter gives rise to lattice fringes. The thin tunnel barrier and superconducting layer are hard to distinguish from the underlying silicon oxide substrate, although are evident in EELS (not shown). Similarly, since there is little difference in atomic number between Co and Cu, the Cu insertion layer is difficult to discern whilst the Au insertion layer appears as a distinct dark band that appears to be continuous. For the Au layer, contrast is already sufficient to conclude that it has a thickness of order 4~nm, although the precise boundaries are hard to distinguish. The random orientation of lattice fringes throughout the Co layers indicate polycrystallinity, without obvious texturing. The interfaces between successive layers appear flat to within one or two nanometres.

\begin{figure}[tbp]
  \begin{indented}\item[]
  \includegraphics[width=13cm]{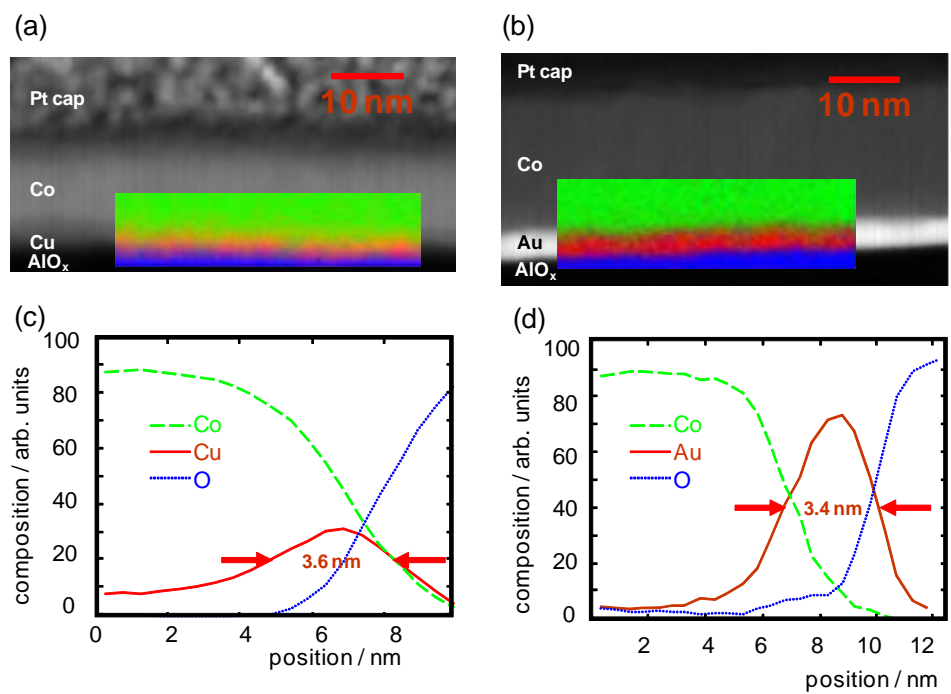}
  \end{indented}
  \caption{STEM and EELS analysis of the SIF devices of figure \ref{tem_sections}. Upper panels (a,b) show dark-field STEM images of the junctions, with (inset) false-coloured compositional profiles derived from EELS spectrum imaging, illustrating Cu (a) or Au (b) in red, Co in green and oxygen in blue. The lower panels (c,d) are projections of the composition showing the vertical distribution of elements across the junction and indicating a discontinuous Cu insertion layer but continuous Au layer. Although the nominal thicknesses differ by a factor of two (1.6~nm for Cu and 3.2~nm for Au) the projected thicknesses are similar. These facts can be reconciled by noting that the Cu layer is discontinuous. Whilst the peak for Cu is much broader than the nominal thickness would imply, it is also much less intense than the equivalent peak for the Au. The total amount of material is therefore consistent with the nominal thickness.}
  \label{stem_eels}
\end{figure}

A combination of STEM imaging and EELS analysis was used to determine the thickness and continuity of the insertion layers. Representative STEM images and EELS data sets collected from the two samples described above are presented in figure \ref{stem_eels}. These data sets are intended to illustrate the wider set of samples considered in this study and to demonstrate our methodology: the figure illustrates an analysis of both Cu and Au layers and also of both continuous and discontinuous layers. The upper panels are black and white dark-field STEM images, in which heavier elements (higher atomic number) appear bright. Superimposed on each image is a small false-coloured panel indicating the region within which EELS spectrum images were acquired.  The colour in each panel derives from the relative proportions of Cu or Au (red), Co (green) and oxygen (blue) at each pixel. Projections (onto the sample growth direction) of the fitted compositional variations are presented in the lower panels, which plot the relative intensity of Co, Cu or Au, and O, moving the across the spectrum image region, perpendicular to the layers and from the Co layer into the barrier and superconductor layer. The barrier is indicated by the presence of oxygen and it should be noted that the intensity calibration is arbitrary since not all elements present were fitted. Note, also, that the projections have not been adjusted for the slight interfacial roughness and vertical shift in layer position, since the trends are clear. Each plot is annotated by the full width at half maximum (FWHM) of the insertion layer signal, which gives an indication of layer thickness.

Both samples show clear Co-rich, Cu (or Au)-rich, and O-rich bands, denoting the ferromagnetic, insertion, and barrier layers, respectively. In the case of the Au sample---figures \ref{stem_eels}(b,d)---has a continuous bright band in the STEM image and continuous Au (red) band in the spectrum image. The FWHM of the STEM intensity peak is measured to be 3.7~nm whilst the intensity of the Au band in the spectrum image has a FWHM of 3.4~nm; thus, we take the Au layer thickness to be 3.6~nm. What is important in figure \ref{stem_eels}(d) is that the Co signal clearly drops before the Au signal, confirming the continuity of the Au layer and proving there to be little or no Co in contact with the barrier. This analysis is important since the TEM and STEM images are essentially two-dimensional projections of around 70~nm of material, so that discrete, discontinuous regions of one material can appear to overlap to form a continuous band: interpreting the STEM image intensity alone is insufficient. This measured thickness is rather close to our assumed nominal thickness of 3.2 nm. In contrast, in the Cu sample---figures \ref{stem_eels}(a,c)---the Cu layer is not continuous: the false-colour panel shows distinct Cu (red) clusters with intermixed Co (green) regions whilst the projected trends clearly indicate the Co and Cu signals to diminish simultaneously, as the O signal rises at the barrier. EELS therefore demonstrates an overlap of copper and oxide but a more detailed analysis of the Cu L$_{2,3}$ edges reveals a lack of the distinctive, sharp `white line' features that would be characteristic of substantial oxide formation. The ‘layer thickness’, which in this case is a measure of the height of discrete Cu nanoparticles above the barrier, is measured to be 3.6~nm. Whilst this is substantially higher than the assumed nominal thickness, that value of 1.6~nm remains a good estimate of the average Cu layer thickness. Given the good match on both cases, we shall discuss our data below in terms of the nominal thicknesses of Cu and Au that we deposited.

Other data sets were analysed in a similar manner and are consistent with a Volmer-Weber growth mode for both Cu and Au, whereby the noble metals form discrete islands or nanoparticles during deposition rather than completely wetting the oxide substrate. Volmer-Weber growth is expected from a surface energy argument, albeit modulated by the non-equilibrium nature of sputter deposition and the restricted kinetics of adatom diffusion on the rough oxide substrate. We collect together some representative data on surface and interface energies $\gamma$ from the literature in table \ref{gammas}. We can see that both the Cu-AlO$_x$ and Au-AlO$_x$ systems satisfy the criterion $\gamma_\mathrm{M} < \gamma_{\mathrm{AlO}_x} + \gamma_{\mathrm{M-AlO}_x}$ for the Volmer-Weber growth mode \cite{venablesbook}, that is non-wetting island growth, at least under thermodynamic conditions. Improved film continuity can be expected due to the out-of-equilibrium nature of sputter growth. The similar surface energetics and identical sputtering conditions for the growth of the Cu and Au layers means we can expect similar growth of both the Cu and Au layers on AlO$_x$. Indeed, these and other EELS data sets are consistent with nanoparticles growing up to a size of order 3~nm before additional material fills in the connecting spaces. This conclusion is similar for both Cu and Au insertion layers and is an agreement with our earlier atomic force microscopy study, which also revealed a nanoparticulate surface \cite{gabureac2008}. Thus, our EELS measurements indicate that both Cu and Au grow as discontinuous layers until a thickness of $\sim$3 nm is achieved.

\begin{table}[btp]
  \caption{\label{gammas}Surface energy $\gamma_\mathrm{M}$ of Cu and Au \cite{deboerbook}, of alumina $\gamma_{\mathrm{AlO}_x}$ \cite{GuzmanCastillo200353}, and interface energy $\gamma_{\mathrm{M-AlO}_x}$ with AlO$_x$ of the two noble metals \cite{pilliar1967}.}
  \begin{indented}
  \item[]\begin{tabular}{@{}llll}
  \br
   & Cu & Au & AlO$_x$ \\
  \mr
  $\gamma_\mathrm{M}$, $\gamma_{\mathrm{AlO}_x}$ (Jm$^{-2}$) & 1.83 & 1.50 & 1.52 \\
  $\gamma_{\mathrm{M-AlO}_x}$ (Jm$^{-2}$) & 1.93 & 1.73 & ---\\
  \br
  \end{tabular}
  \end{indented}
\end{table}

The implication of our TEM data is that the interface between oxide and insertion layer is inhomogeneous. In the early stages of growth a decreasing fraction of bare oxide is retained while metal nanoparticles grow and merge. Thus, there are two distinct electronic pathways to consider for spin-dependent transport across the SIF device: that directly from oxide to Co and that mediated by the Cu (Au) nanoparticles. Only when those particles exceed a height of order 3~nm is the oxide completely covered and thereafter the film can be expected to increase in thickness more uniformly. In this case all tunnelling must take place mediated by the noble metal layer.

\subsection{Interface-doped junctions}

\begin{figure}[tbp]
  \begin{indented}\item[]
  \includegraphics[width=7.5cm]{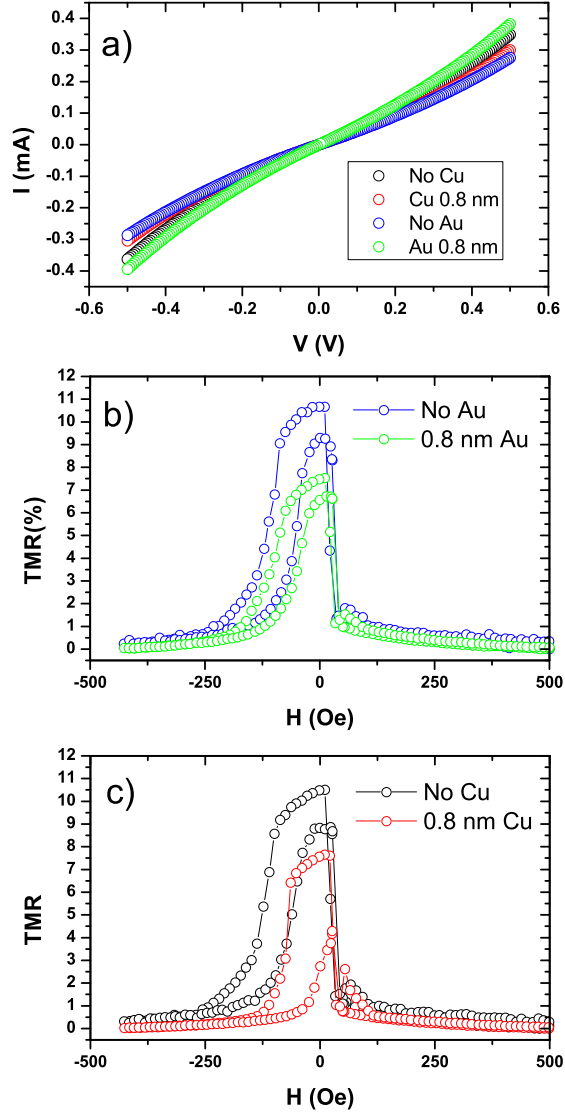}
  \end{indented}
  \caption{Room temperature transport properties of F-I-F junctions. (a) $I$-$V$ characteristics. (b) TMR of a junction with an 0.8~nm thick Au layer, compared with an undoped control sample. (These data were previously shown in reference \cite{gabureac2008}.) (c) TMR of a junction with an 0.8~nm thick Cu layer, compared with an undoped control sample.}
  \label{ivfif}
\end{figure}

Next we discuss the effects of interface-doping an FIF structure on the magnetotransport. In figure \ref{ivfif} we show room temperature $I$-$V$ characteristics and TMR data for Au and Cu interface doped MTJs, along with undoped control samples from the same growth runs. All the data were measured at room temperature. In the two doped samples there is an 0.8~nm thick layer of noble metal, Au or Cu, introduced at the interface: thick enough only to form a layer of noble metal nanoparticles at the interface. The $I$-$V$ characteristics in panel (a) show a small variation but are overall rather similar: the variation is not correlated with any junction parameter. The junction resistances are in the few k$\Omega$ range, yielding resistance area products of several M$\Omega \mu$m$^2$. The introduction of the noble metal layers reduces the TMR ratio from $\sim 10.5$~\% to $\sim 7.5$~\% in both cases, as can be seen in panels (b) and (c). The switching properties of the junctions are not greatly affected by the doping, showing that the loss of TMR is due to the change in the effective TSP of the electrodes.

\begin{figure}[tbp]
  \begin{indented}\item[]
  \includegraphics[width=7.5cm]{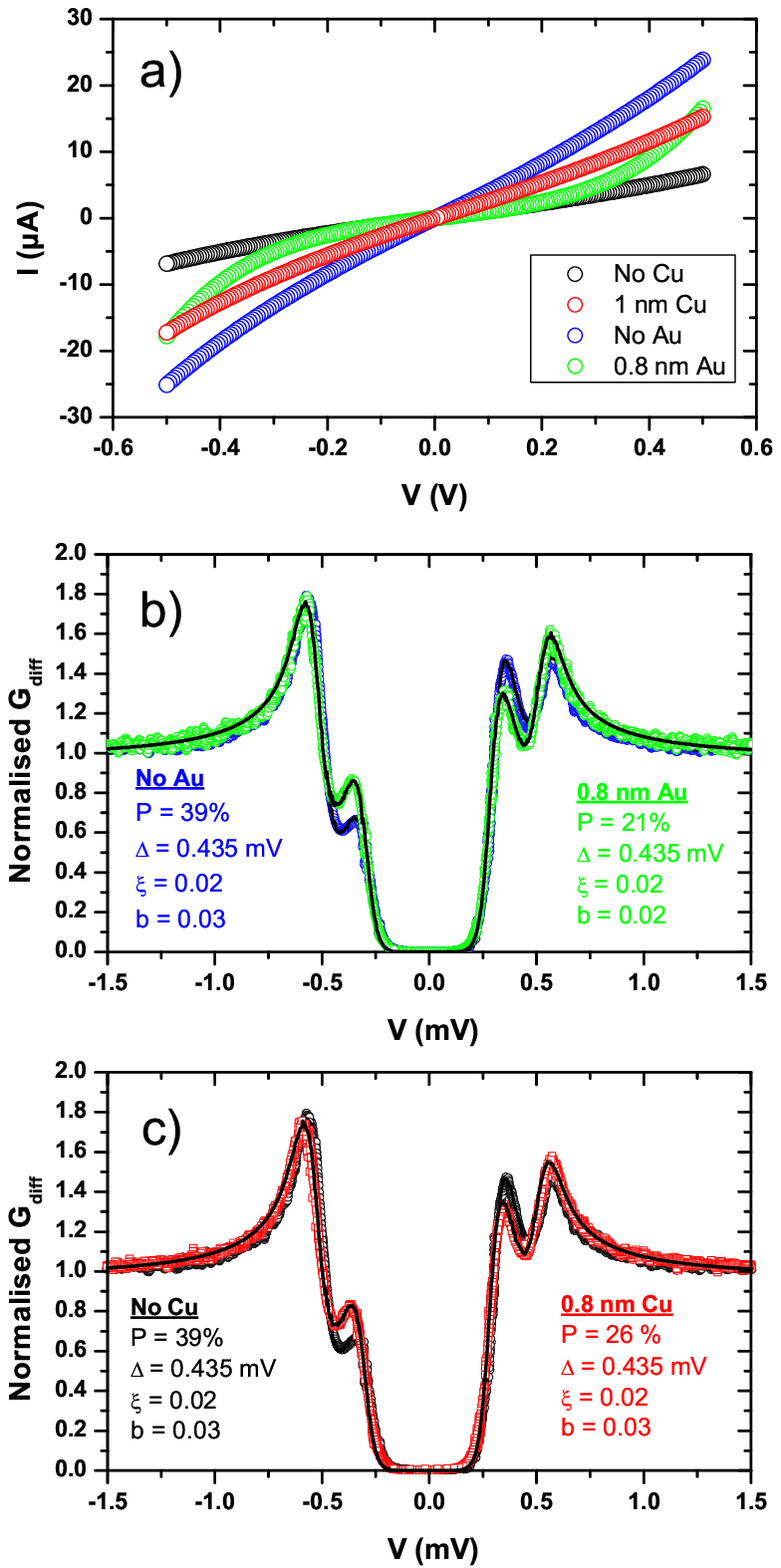}
  \end{indented}
  \caption{Transport properties of S-I-F junctions. (a) $I$-$V$ characteristics. (b) Normalised $G_{\rm diff}$ of a junction with an 0.8~nm thick Au layer, compared with an undoped control sample. (These data and fits were previously shown in reference \cite{gabureac2008}.) (c) Normalised $G_{\rm diff}$ of a junction with an 0.8~nm thick Cu layer, compared with an undoped control sample. The black solid lines are fits of the AORT model to the data, as described in the text. The parameters returned by the fit are inset in each panel of the figure.}
  \label{ivsif}
\end{figure}

The TSP can be measured directly by the Meservey-Tedrow method \cite{meserveytedrow}. In figure \ref{ivsif} we show room temperature $I$-$V$ characteristics, in panel (a), and $G_\mathrm{diff} = dI/dV$ data for Au and Cu interface doped SIF junctions in panels (b) and (c), along with control samples from the same growth runs. Again, in each doped sample there is a layer of noble metal introduced at the interface: nominally 0.8~nm thick for both Au and Cu. There is a little more variation here in the $I$-$V$ characteristics than in the previous case, but again it is not obviously correlated with any specific junction property. The junction resistances range between $\sim 25$ k$\Omega$ and $\sim 90$ k$\Omega$, putting the resistance area products into the 100 M$\Omega \mu$m$^2$ range. This is much higher than for the MTJs, suggesting that we have, in fact, oxidised a little more deeply into the Al$_{96}$Si$_4$ layer than the nominal 1.3~nm depth.

The four peaks corresponding to the Zeeman split edges of the BCS (Bardeen-Cooper-Schrieffer) gap in the Al$_{96}$Si$_4$ electrode are clearly visible in the $G_{\rm diff}$ data in both panels (b) and (c) of figure \ref{ivsif}, which are shown normalized to the quasiparticle conductance beyond the gap. The data are fitted by the model of Alexander, Orlando, Rainer, and Tedrow (AORT) \cite{aort,worledge}. In this model there are four input parameters: the BCS gap $\Delta$, the spin-orbit parameter $b$ (which accounts for all the mechanisms that mix spin up and down without destroying the Cooper pairs), the spin-flip or depairing parameter $\xi$ (which includes all mechanisms that break the time-reversal symmetry and hence break Cooper pairs), and the TSP itself, $P$. The fits to the data are shown in the plot as solid lines, along with the values of these parameters. All four samples are well described by similar values for $\Delta$, $b$, and $\xi$, which are typical for an Al electrode doped with Si \cite{kaiser2004}. The TSP of both of our undoped samples is $P = 39$~\%, a typical value for a transition metal \cite{meserveytedrow}. This is roughly halved, to 21~\%, on the introduction of the 0.8~nm Au interface layer, and is slightly higher, 25.5~\%, in the sample with the Cu interface layer.

%Fig. \ref{resistance} shows data for the resistance of the interface doped junctions. {\bf More details needed here: I don't quite understand the difference between all the various resistances that are plotted. Need we do some Simmons fitting of the I-V characteristics to see how barrier parameters vary?}
%
%\begin{figure}[tbp]
%  \onefigure[width=8cm]{resistance.eps}
%  \caption{(Colour online) Transport properties of junctions doped with Au and Cu.} \label{resistance}
%\end{figure}

\begin{figure}[tbp]
  \begin{indented}\item[]
  \includegraphics[width=12.5cm]{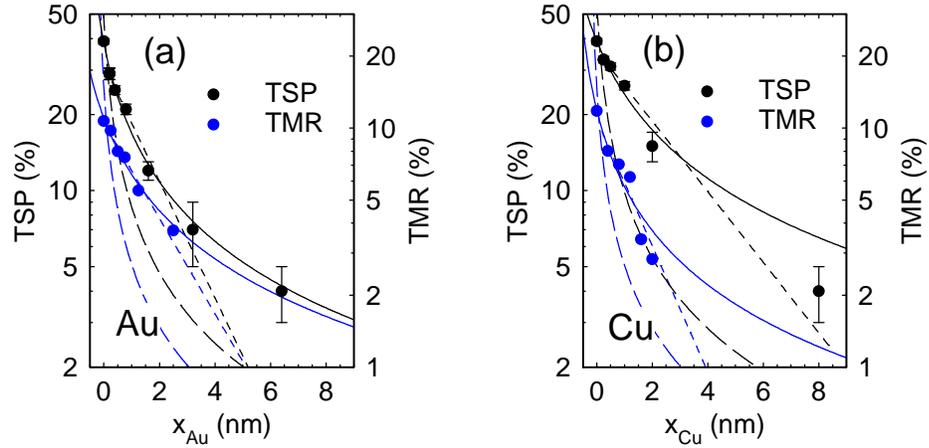}
  \end{indented}
  \caption{(Colour online) Transport properties of junctions doped at the Co/AlO$_x$ interface with (a) Au and (b) Cu. The TSP is plotted on the left-hand ordinate in each graph, the TMR on the right-hand one. The dotted lines are fits of an exponential decay to the data, which appear as straight lines on these semi-log plots. The dashed lines are a fit of the model described in reference \cite{moodera1989} with $a = 0.223$~nm, the value for the atomic size of Co. The solid lines are fits with $a$ as a variable parameter. (The data, but not the analysis, in panel (a) were previously shown in reference \cite{gabureac2008}.)} \label{interfacedecay}
\end{figure}

Having looked at these examples for samples with and without thin (discontinuous) interface layers, we now move on to discuss the systematic way in which the TMR and TSP  presented at the barrier varies as the thickness, $x$, of these interfacial layers: in figure \ref{interfacedecay} we show data for a variety of values of $x$ for both Au and Cu, determined using the techniques described above, including the data points obtained from the measurements depicted in figures \ref{ivfif} and \ref{ivsif}. The data for the junctions with Au interface layers are plotted in panel (a). For undoped samples we obtain a Meservey-Tedrow TSP of $39 \pm 1$~\%, whilst the room temperature TMR of 10.7~\% implies, through the Julliere formula \cite{julliere}, that the geometric mean of the polarisations of the two electrodes is 22.5~\%. This discrepancy can be explained partly through elevated temperature at which the TMR measurements are made \cite{hindmarch2005}, and perhaps also partly through suboptimal oxidation of the barrier reducing the TSP of the bottom electrode in the FIF stack and hence the geometric mean TSP: we have not performed any anneal, which typically improves such a situation. Finite values of TSP extend to values of $x$ well beyond the point at which the TEM studies show that the noble metal layers become continuous.

% Relevant form of Julliere is P = +/- sqrt (TMR / 2 + TMR)

There is a small drop in both TMR and TSP when the Au is first introduced, but, remarkably, an easily measurable TMR and TSP of a few \% is still present for Au thicknesses exceeding 2~nm. The TEM has shown that the  Au layer is not fully continuous by this point. Nevertheless, the layer should cover more than the 50 \% of the surface of the barrier that the reduction in TMR implies if the suppression is solely due to contact areas between the Co and the AlO$_x$. More compelling as evidence for long-ranged magnetic proximity effects, there is a finite and substantial TSP that can be measured for even several nanometres of Au between the barrier and the Co, \textit{i.e.} well into the regime where the TEM studies tell us that the noble metal layer is completely continuous. In order to quantify the distance that the spin-polarisation propagates through the Au, the data were fitted by an exponential decay, $\propto \exp(-x_\mathrm{Au} / \lambda)$, which are shown as dotted lines. These fits yielded $1/e$ decay lengths $\lambda$ for both the TSP and TMR that are given in table \ref{lambdas}. An equivalent set of data for junctions with Cu interface layers is plotted in panel (b) of figure \ref{interfacedecay}. Again we obtain reasonably good exponential fits to most of the data, the exception being the TSP at very large interface layer thickness.

The $1/e$ decay lengths are all of order of 2~nm, with the exception of $\lambda_\mathrm{TSP}$ for Cu, which is somewhat longer at 3.2~nm. Nevertheless, in all cases it is remarkable that the decay lengths are so long. For Au the two decay lengths are roughly the same, with $\lambda_\mathrm{TMR} \approx \lambda_\mathrm{TSP}$. However, for Cu $\lambda_\mathrm{TMR}$ is roughly half as long as $\lambda_\mathrm{TSP}$, which is probably to be attributed to the TSP measurement being made at a much lower temperature. The stronger spin-orbit interaction in Au may limit this low temperature lengthscale. Furthermore, the $\lambda$ values given here may be regarded as lower limits on the true values, since they arise from fits to all the data points, including those for small $x$ where the inserted noble metal layers are not fully continuous. The values for TSP and TMR at low $x$ are overestimates, since there is contribution from those parts of sample where the Co is in direct contact with the tunnel barrier. If this were accounted for (difficult to do in practice with our limited number of data points) the changes in TMR and TSP would be less steep and the $\lambda$ values even larger.

\begin{table}[btp]
  \caption{\label{lambdas} $1/e$ decay lengths $\lambda$ and exchange scattering lengths $a$ arising from the fits to the TSP and TMR data shown in figure \ref{interfacedecay} for interface doped samples.}
  \begin{indented}
  \item[]
  \begin{tabular}{@{}lll}
  \br
   & Au & Cu  \\
  \mr
  $\lambda_\mathrm{TSP}$ (nm) & $1.9 \pm 0.3$ & $3.2 \pm 0.4$ \\
  $\lambda_\mathrm{TMR}$ (nm) & $2.2 \pm 0.3$ & $1.7 \pm 0.4$ \\
  \mr
  $a_\mathrm{TSP}$ (nm) & $0.79 \pm 0.07$ & $1.6 \pm 0.3$ \\
  $a_\mathrm{TMR}$ (nm) & $1.4 \pm 0.2$    & $9.0 \pm 0.2$  \\
  \br
  \end{tabular}
  \end{indented}
\end{table}

Whilst useful for estimating lengthscales, these exponential fits are purely phenomenological. They also seem to underestimate the TSP that will be observed in the cases where we have the thickest Cu or Au interface layers, suggesting a power law decay, rather than an exponential one, is appropriate, at least for large $x$. Moodera, Taylor, and Meservey performed a very similar experiment to ours on Au layers between an alumina barrier and an Fe electrode \cite{moodera1989}, they also observed finite TSP out to several nanometres. They developed a simple model to interpret their data, in which the electrons in the Au that are close to the Fermi level are spin-polarised by exchange scattering from the Fe atoms at the interface between the two metals. They derived a formula predicting that the TSP should be proportional to $a/(x + a)$, where $a$ is the cube root of the atomic volume of the magnetic atom species, which fitted their data well using the value of $a = 0.228$~nm that is appropriate for for Fe. An attempt to fit this functional form (with the constraint $a = 0.223$~nm, the value for Co) to each of the various sets of data is also shown in figure \ref{interfacedecay} using dashed lines. The fit is rather poor. However, it can be improved markedly by allowing $a$ to be a variable fitting parameter, with the results of doing so shown as solid lines in figure \ref{interfacedecay}. The results returned for $a$ are also given in table \ref{lambdas}. In all cases the uncertainties in $a$ are smaller than those in $\lambda$, showing that this model describes the data somewhat better than the phenomenological exponential decay. The values for $a$ are all at least a substantial fraction of a nanometre. Again, the contribution at low $x$ from Co directly in contact with the barrier means that these values should be regarded as lower limits on $a$. This indicates that this exchange scattering mechanism involves the first few atomic layers of Co, rather than just the first one, as in the case of Fe. This may be because the spin-$\uparrow$ band of Co is a good match for the band structure of Cu and Au \cite{papaconstantopoulosbook}, giving a low interface scattering rate for electrons of this spin, permitting them to easily enter the Co layer and scatter within its bulk. There is no such good match between Fe and Au, meaning that there is strong interface scattering and only the first atomic layer of Fe plays a role, consistent with the report of Moodera \textit{et al.} \cite{moodera1989}.

\subsection{$\delta$-doped junctions}

\begin{figure}[tbp]
  \begin{indented}\item[]
  \includegraphics[width=12.5cm]{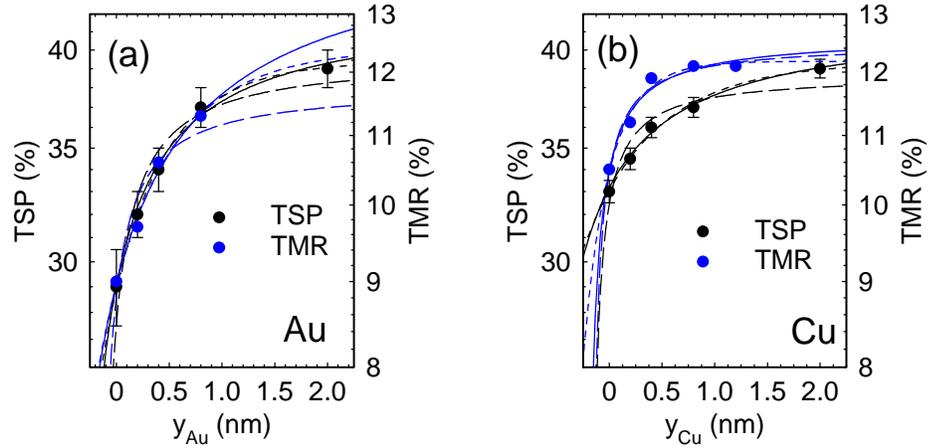}
  \end{indented}
  \caption{Transport properties of junctions $\delta$-doped within the Co electrode with (a) Au, and (b) Cu. The TSP is plotted on the left-hand ordinate in each graph, the TMR on the right-hand one. The dotted lines are fits of an exponential rise to the data. The dashed lines are a fit of the model described in reference \cite{moodera1989} with the modifications described in the text, with $b$ set to the value for the atomic size for Au ($b = 0.257$~nm) or Cu ($b = 0.227$~nm). The solid lines are fits with $b$ as a variable parameter. (The data, but not the analysis, in panel (a) were previously shown in reference \cite{gabureac2008}.) }
  \label{bulkdecay}
\end{figure}

Having studied the magnetic proximitisation of Cu and Au by Co, it is interesting to study whether any inverse proximity effects are also present. This was done using the $\delta$-layer technique originally established by Parkin to study giant magnetoresistance (GMR) \cite{parkin1993}, and more recently used by some of us to study GMR \cite{marrows2001} and exchange bias \cite{ali2008}. In this method, an unltrathin $\delta$-layer of impurity, in this case Cu or Au, is introduced into the sample at a point in the Co layer that is a distance $y$ from the interface with the barrier, as depicted in figure \ref{stacks} (b) and (d). In figure \ref{bulkdecay} we show the results of an equivalent set of measurements for the $\delta$-doped samples to those for the interface-doped samples presented in figure \ref{bulkdecay}. Here a $\delta$-layer of Au or Cu that is $\sim 0.1$~nm thick was inserted into the Co layer a distance $y$ from the interface with the barrier. The introduction of both Au and Cu at the interface ($y=0$) suppresses the TMR and TSP in a manner that is consistent with the results presented in figure \ref{interfacedecay} for $x \approx 0.1$~nm. As the noble metal is moved into the Co layer, the TSP, and hence TMR, recovers as the Co-like spin-polarised band structure is restored at the interface. This is fully restored to the value obtained without any doping once the $\delta$-layer is 1-2~nm away from the tunnel barrier.

In order to properly quantify this recovery, we again fit the data to various models. Once more, the dotted lines show the results of a best fit of a phenomenological exponential rise. The $1/e$ lengthscales $\lambda$ for these fits are given in table \ref{lambdasbulk}. The values are all rather short, being only $\sim 0.6$-$0.7$~nm, with the exception of $\lambda_\mathrm{TMR}$ for Cu, which is even shorter, only $\sim 0.3$~nm. Evidently, whatever physics is at play, the full spin-polarisation presented to the tunnel barrier is quickly restored.

\begin{table}[btp]
  \caption{\label{lambdasbulk} $1/e$ decay lengths $\lambda$ and exchange scattering lengths $a$ arising from the fits to the TSP and TMR data shown in figure \ref{bulkdecay} for bulk $\delta$-doped samples.}
  \begin{indented}
  \item[]
  \begin{tabular}{@{}lll}
  \br
   & Au & Cu  \\
  \mr
  $\lambda_\mathrm{TSP}$ (nm) & $0.57 \pm 0.04$ & $0.7 \pm 0.1$ \\
  $\lambda_\mathrm{TMR}$ (nm) & $0.7 \pm 0.4$ & $0.28 \pm 0.08$ \\
  \mr
  $a_\mathrm{TSP}$ (nm) & $0.63 \pm 0.01$ & $0.96 \pm 0.01$ \\
  $a_\mathrm{TMR}$ (nm) & $1.1 \pm 0.9$    & $0.3 \pm 0.2$     \\
  \br
  \end{tabular}
  \end{indented}
\end{table}

Theory is less well-developed for this configuration of sample. It is possible to argue by analogy with the model of Moodera \textit{et al.} in reference \cite{moodera1989} that the TSP ought to depend on the thickness of the Co layer $y$ according to $y/(y + b)$, where $b$ is now a suitable scattering length for the noble metal species. (A caveat that needs to be borne in mind is that the noble metal is now a thin $\delta$-layer rather than a semi-infinite layer which the model assumes.) Constraining $a$ to take the values appropriate to the cube root of the atomic volume ($b = 0.257$~nm for Au and $b = 0.227$~nm for Cu) gives rise to the fits shown as dashed lines in figure \ref{bulkdecay} which are rather poor in most cases --- the exception is for the measurements of TMR with Cu $\delta$-layers, where the good fit may just be fortuitous. For the sake of completeness, we also show fits where $b$ is allowed to vary as a fitting parameter: these are the solid lines in figure \ref{bulkdecay}. The fits to the data are good, and the values for $a$ returned by them are reported in table \ref{lambdasbulk}. In the regions where data points exist, it is difficult to distinguish between this model and the exponential fits - this is unlike the case of interface doping where the exchange scattering model of reference \cite{moodera1989} is palpably better at describing the data. Where the fitted values for $b$ are all substantially longer than the $\sim 0.1$~nm thickness of the $\delta$-layer, the value of this model is more doubtful. However, it is clear that for a layer of Co next to the tunnel barrier only a very few atomic diameters thickness is required to restore the full TSP, regardless of the underlying physics.

\section{Conclusion}

To conclude, we have shown that measurable TSP, leading to a finite TMR ratio, can penetrate a few nm into both Au and Cu to present itself at the tunnel barrier, even when the noble metal insertion layer has been confirmed by TEM to be fully continuous. This is our main result, which is purely experimental, being independent of the model used to fit the data, and implies a proximitisation of the transport-active electrons in the noble metal by the exchange field in the Co. The full TSP is restored when the noble metal layer is withdrawn into the Co layer by only about 1~nm, indicating the short range of the associated inverse proximity effect.

However, the substantial values of TSP and TMR that we observe appear to contradict various XMCD studies that only find very small magnetic moments proximitised layers. An important difference to note is that the XMCD method is primarily sensitive to the $d$ electrons, whereas transport measurements (especially tunnelling \cite{stearns1977}) are more sensitive the the $s$-like electrons, and in particular those at the Fermi level. This point was made by Giordano and Pennington, who measured weak localisation effects in Au nanowires in contact with an Fe lead and found evidence for spin-polarisation effects extending $\sim 1~\mu$m into the Au at 4 K \cite{giordano1992}. This is to be contrasted with hot electron photoemission results, such as those of Pierce and Siegmann, who measured a 1~nm decay length through Cu, where electrons further from the Fermi level are probed \cite{pierce1974}. The XMCD results cited in the introduction \cite{samant1994,bartolome2008,huttel2008,abes2009,springell2008,maccherozzi2008} also concern (primarily $d$-like) electrons over a wider energy window around the Fermi level. Our results here were obtained at low biases---10~mV for TMR and $<1$~mV for the TSP---and so are very sensitive to what happens at the Fermi level. On this basis we can thus account for the differences between those results and the ones we report here.

The issue of what is the relevant lengthscale was addressed theoretically by Zhang and Levy, who calculated the effects of inserting an additional layer into a magnetic tunnel junction within a quasi-ballistic picture \cite{zhang1998}. They found that lengthscales comparable to those that we report here can be be interpreted as being those over which the wavefunctions of electrons tunnelling from neighbouring points on the barrier remain coherent with one another, setting the decay length for the TSP or TMR. (This is distinct from the lengthscale over which the dephasing of wavefunctions takes place, which can be very long in Cu and Au, particularly at low temperatures \cite{pierre2003}.) In the case where the spin-asymmetry of the interface resistance is high (which will lead to high giant magnetoresistance), as it is in the case of Co/Cu or Co/Au, that spin-asymmetry can be maintained over that decay length. A related argument was made by Parkin \cite{parkinspacerpatent}. Whilst it is difficult to quantify the resulting decay length, since it depends in a detailed way on the interface roughnesses in the hybrid MTJ stack \cite{zhang1998}, these authors comment on unpublished data from Parkin that show decay lengths of about 4~nm in Cu and 3~nm in Ag (subsequent publication shows a decay length of 3.69~nm in Co-based MTJs with Cu deposited on top of the tunnel barrier, and 0.84~nm with Au layers \cite{parkinspacerpatent}). These are distances that, like the ones we report here, comfortably exceed the width of a single plane of atoms that is often taken to be depth of a tunnelling electrode that contributes to a tunnel current. Although framed for FIF tunnel junctions, their arguments apply with equal for SIF structures as well. Thus we can conclude that the magnetic proximity effects we observe here are mediated by the spin-polarised $s$-like electrons that can maintain their coherence with each other over a non-magnetic noble metal spacer that is several nanometres thick. The fact that TMR and TSP is restored so quickly when the noble metal is withdrawn only a short distance into the Co further shows how robust a phenomenon spin-polarisation is in these electronically well-matched materials systems.

\ack
We would like to thank A.~T.~Hindmarch for help with the coding of the AORT model, S.~Rajauria for assistance with the Meservey-Tedrow measurements, and the EU (via project NMP2-CT-2003-505587 ``SFINx'') and the UK Engineering and Physical Sciences Research Council (grant EP/E016413/1) for funding.

% Create the reference section using BibTeX

%\bibliographystyle{iopart-num}
%\bibliography{impuritytunnelling}

\begin{thebibliography}{10}
\expandafter\ifx\csname url\endcsname\relax
  \def\url#1{{\tt #1}}\fi
\expandafter\ifx\csname urlprefix\endcsname\relax\def\urlprefix{URL }\fi
\providecommand{\eprint}[2][]{\url{#2}}
% Bibliography created with iopart-num v2.1
% /biblio/bibtex/contrib/iopart-num

\bibitem{michael_tinkham_introduction_1996}
Tinkham M 1996 {\em Introduction to Superconductivity\/} 2nd ed (New York:
  McGraw Hill)

\bibitem{samant1994}
Samant M~G, St\"{o}hr J, Parkin S~S~P, Held G~A, Hermsmeier B~D, Herman F, {van
  Schilfgaarde} M, Duda L~C, Mancini D~C, Wassdahl N and Nakajima R 1994 {\em
  Phys. Rev. Lett.\/} {\bf 72} 1112

\bibitem{bartolome2008}
Bartolom\'{e} J, García L~M, Bartolom\'{e} F, Luis F, L\'{o}pez-Ruiz R, Petroff
  F, Deranlot C, Wilhelm F, Rogalev A, Bencok P, Brookes N~B, Ruiz L and
  Gonz\'{a}lez-Calbet J~M 2008 {\em Phys. Rev. B\/} {\bf 77} 184420

\bibitem{huttel2008}
Huttel Y, Clavero C, {van der Laan} G, Bencok P, Johal T~K, Claydon J~S,
  Armelles G and Cebollada A 2008 {\em Phys. Rev. B\/} {\bf 77} 064411

\bibitem{abes2009}
Abes M, Atkinson D, Tanner B~K, Charlton T, Langridge S, Hase T~P~A, Ali M,
  Marrows C~H, Neudert A, Hicken R~J, Mirone A and Arena D 2009 {\em J. Appl.
  Phys.\/} {\bf 105} 07C703

\bibitem{springell2008}
Springell R, Wilhelm F, Rogalev A, Stirling W~G, Ward R~C~C, Wells M~R,
  Langridge S, Zochowski S~W and Lander G 2008 {\em Phys. Rev. B\/} {\bf 77}
  064423

\bibitem{maccherozzi2008}
Maccherozzi F, Sperl M, Panaccione G, Min\'{a}r J, Polesya S, Ebert H,
  Wurstbauer U, Hochstrasser M, Rossi G, Woltersdorf G, Wegscheider W and Back
  C~H 2008 {\em Phys. Rev. Lett.\/} {\bf 101} 267201

\bibitem{mark2009}
Mark S, Gould C, Pappert K, Wenisch J, Brunner K, Schmidt G and Molenkamp L~W
  2009 {\em Phys. Rev. Lett\/} {\bf 103} 017204

\bibitem{olejnik2010}
Olejnik K, Wadley P, Haigh J~A, Edmonds K~W, Campion R~P, Rushforth A~W,
  Gallagher B~L, Foxon C~T, Jungwirth T, Wunderlich J, Dhesi S~S, Cavill S~A,
  {van der Laan} G and Arenholz E 2010 {\em Phys. Rev. B\/} {\bf 81} 104402

\bibitem{nie2013}
Nie S~H, Chin Y~Y, Liu W~Q, Tung J~C, Lu J, Lin H~J, Guo G~Y, Meng K~K, Chen L,
  Zhu L~J, Pan D, Chen C~T, Xu Y~B, Yan W~S and Zhao J~H 2013 {\em Phys. Rev.
  Lett.\/} {\bf 111} 027203

\bibitem{stohr1999}
St\"{o}hr J 1999 {\em J. Magn. Magn. Mater.\/} {\bf 200} 470

\bibitem{meserveytedrow}
Meservey R and Tedrow P~M 1994 {\em Phys. Rep.\/} {\bf 238} 173

\bibitem{julliere}
Julliere M 1975 {\em Phys. Lett. A\/} {\bf 54} 225

\bibitem{baderandparkin}
Bader S and Parkin S 2010 {\em Annual Review of Condensed Matter Physics\/}
  {\bf 1} 71

\bibitem{moodera1995}
Moodera J~S, Kinder L~R, Wong T~M and Meservey R 1995 {\em Phys. Rev. Lett\/}
  {\bf 74} 3273

\bibitem{yuasanatmat2004}
Yuasa S, Nagahama T, Fukushima A, Suzuki Y and Ando K 2004 {\em Nature
  Materials\/} {\bf 3} 868

\bibitem{parkinnatmat2004}
Parkin S~S~P, Kaiser C, Panchula A, Rice P~M, Hughes B, Samant M and Yang S~H
  2004 {\em Nature Materials\/} {\bf 3} 862

\bibitem{slonczewski1989}
Slonczewski J~C 1989 {\em Phys. Rev. B\/} {\bf 39} 6995

\bibitem{tedrowmeservey}
Tedrow P~M and Meservey R 1975 {\em Solid State Comm.\/} {\bf 16} 71

\bibitem{marrows2001}
Marrows C~H and Hickey B~J 2001 {\em Phys. Rev. B\/} {\bf 63} 220405

\bibitem{parkin1993}
Parkin S~S~P 1993 {\em Phys. Rev. Lett.\/} {\bf 71} 1641

\bibitem{stanley2000}
Stanley F~E, Marrows C~H and Hickey B~J 2000 {\em J. Appl. Phys.\/} {\bf 87}
  4864

\bibitem{moodera1999}
Moodera J~S, Nowak J, Kinder L~R, Tedrow P~M, {van de Veerdonk} R~J~M, Smits
  B~A, {van Kampen} M, Swagten H~J~M and de~Jonge W~J~M 1999 {\em Phys. Rev.
  Lett.\/} {\bf 83} 3029

\bibitem{sun1999}
Sun J~J and Freitas P~P 1999 {\em J. Appl. Phys.\/} {\bf 85} 5264

\bibitem{leclair2000}
LeClair P, Swagten H~J~M, Kohlhepp J~T, {van de Veerdonk} R~J~M and de~Jonge
  W~J~M 2000 {\em Phys. Rev. Lett.\/} {\bf 84} 2933

\bibitem{yuasa2002}
Yuasa S, Nagahama T and Suzuki Y 2002 {\em Science\/} {\bf 297} 234

\bibitem{samant2004}
Samant M~G and Parkin S~S~P 2004 {\em Vacuum\/} {\bf 74} 705

\bibitem{leclair2001cr}
LeClair P, Kohlhepp J~T, Swagten H~J~M and de~Jonge W~J~M 2001 {\em Phys. Rev.
  Lett.\/} {\bf 86} 1066

\bibitem{leclair2001}
LeClair P, Hoex B, Wieldraaijer H, Kohlhepp J~T, Swagten H~J~M and de~Jonge
  W~J~M 2001 {\em Phys. Rev. B\/} {\bf 64} 100406

\bibitem{nozaki2004}
Nozaki T, Jiang Y, Kaneko Y, Hirohata A, Tezuka N, Sugimoto S and Inomata K
  2004 {\em Phys. Rev. B\/} {\bf 70} 172401

\bibitem{gabureac2008}
Gabureac M~S, Dempsey K~J, Porter N~A, Marrows C~H, Rajauria S and Courtois H
  2008 {\em J. Appl. Phys.\/} {\bf 103} 07A915

\bibitem{moodera1989}
Moodera J~S, Taylor M~E and Meservey R 1989 {\em Phys. Rev. B\/} {\bf 40} 11980

\bibitem{scott2008}
Scott J, Thomas P~J, MacKenzie M, McFadzean S, Wilbrink J, Craven A~J and
  Nicholson W~A~P 2008 {\em Ultramicroscopy\/} {\bf 108} 1586

\bibitem{hunt1991}
Hunt J~A and Williams D~B 1991 {\em Ultramicroscopy\/} {\bf 38} 47

\bibitem{luetkens2003}
Luetkens H, Korecki J, Morenzoni E, Prokscha T, Birke M, Gl\"{u}ckler H,
  Khasanov R, Klauss H~H, \'{S}lezak T, Suter A, Forgan E~M, Niedermayer C and
  Litterst F~J 2003 {\em Phys. Rev. Lett.\/} {\bf 91} 017204

\bibitem{venablesbook}
Venables J~A 2000 {\em Introduction to Surface and Thin Film Processes\/}
  (Cambridge: Cambridge University Press)

\bibitem{deboerbook}
{de Boer} F~R, Boom R, Mattens W~C~M, Miedema A~R and Niessen A~K 1988 {\em
  Cohesion in Metals: Transition Metal Alloys\/} (Amsterdam: North-Holland)

\bibitem{GuzmanCastillo200353}
Guzm\'{a}án-Castillo M~L, Bokhimi X, Rodr\'{\i}guez-Hern\'{a}ndez A,
  Toledo-Antonio A, Hern\'{a}ndez-Beltr\'{a}n F and Fripiat J~J 2003 {\em J.
  Non-Cryst. Solids\/} {\bf 329} 53 -- 56

\bibitem{pilliar1967}
Pilliar R~M and Nutting J 1967 {\em Phil. Mag.\/} {\bf 16} 181--188

\bibitem{aort}
Alexander J~A~X, Orlando T~P, Rainer D and Tedrow P~M 1985 {\em Phys. Rev. B\/}
  {\bf 31} 5811

\bibitem{worledge}
Worledge D~C and Geballe T~H 2000 {\em Phys. Rev. B\/} {\bf 62} 447

\bibitem{kaiser2004}
Kaiser C and Parkin S~S~P 2004 {\em Appl. Phys. Lett.\/} {\bf 84} 3582

\bibitem{hindmarch2005}
Hindmarch A~T, Marrows C~H and Hickey B~J 2005 {\em Phys. Rev. B\/} {\bf 72}
  060406

\bibitem{papaconstantopoulosbook}
Papaconstantopoulos D~A 1986 {\em Handbook Of The Band Structure Of Elemental
  Solids\/} (Berlin: Springer)

\bibitem{ali2008}
Ali M, Marrows C~H and Hickey B~J 2008 {\em Phys. Rev. B\/} {\bf 77}(13) 134401

\bibitem{stearns1977}
Stearns M~B 1977 {\em J. Magn. Magn. Mater.\/} {\bf 5} 167

\bibitem{giordano1992}
Giordano N and Pennington M~A 1992 {\em Phys. Rev. B\/} {\bf 45} 14238--14246

\bibitem{pierce1974}
Pierce D~T and Siegmann H~C 1974 {\em Phys. Rev. B\/} {\bf 9} 4034

\bibitem{zhang1998}
Zhang S and Levy P~M 1998 {\em Phys. Rev. Lett.\/} {\bf 81} 5660

\bibitem{pierre2003}
Pierre F, Gougam A~B, Anthore A, Pothier H, Esteve D and Birge N~O 2003 {\em
  Phys. Rev. B\/} {\bf 68} 085413

\bibitem{parkinspacerpatent}
Parkin S~S~P 1998 Magnetic tunnel junction device with nonferromagnetic
  interface layer for improved magnetic field response {U.S.} patent 5,746,767

\end{thebibliography}

\providecommand{\newblock}{}

\end{document}